\RequirePackage{fix-cm}
\documentclass[twocolumn]{svjour3}          
\smartqed  
\usepackage{graphicx}
\begin{document}

\title{Theoretical possibilities for flat-band superconductivity
}


\author{Hideo Aoki}


\institute{H. Aoki \at
              Dept. of Physics, University of Tokyo, 
Hongo, Tokyo 113-0033, Japan, 
and National Institute of Advanced Industrial Science and Technology (AIST), 
Tsukuba 305-8568, Japan\\
              \email{aoki@phys.s.u-tokyo.ac.jp}           
         }

\date{Received: date / Accepted: date}

\maketitle

\begin{abstract}
One novel arena for designing superconductors with high $T_C$ is the flat-band systems.  A basic idea is that flat bands, arising from quantum mechanical interference, give unique opportunities for enhancing $T_C$ with (i) many pair-scattering channels between the dispersive and flat bands, and (ii) an even more interesting situation when the flat band is topological and highly entangled.  Here we compare two routes, which comprise a multi-band system with a flat band coexisting with dispersive ones, and a one-band case with a portion of the band being flat.  Superconductivity can be induced in both cases when the flat band or portion is ``incipient" (close to, but away from, the Fermi energy).  Differences are, for the multi-band case, we can exploit large entanglement associated with topological states, while for the one-band case a transition between different (d and p) wave pairings can arise.  These hint at some future directions.  
\end{abstract}

\section{Introduction}
\label{intro}
In the long history of studies of correlated electron 
systems, superconductivity in repulsively interacting 
electrons are known to sensitively depend on the underlying band structures.  
Indeed, this is a first point discerning different classes of 
superconductors exemplified by the cuprates, iron-based, organic, etc. 
Then a question is how we can engineer them for (i) 
favouring superconductivity (SC), and (ii) control 
the pairing symmetry, with different symmetries possibly coexisting. 
The thesis of the present paper is that the flat-band systems 
provide an interesting and unique arena for those.

Let us start with a very general question: 
Which is most favourable for SC, one-band systems, or 
multi-band systems (comprising either multi-orbitals or single-orbital)?   
As far as the ordinary (dispersive) bands are concerned, 
Sakakibara et al. have theoretically shown, 
for the case of multi-orbital multi-band systems, that 
the farther the second band (with e.g. $d_{z^2}$ 
orbital character) the higher the $T_C$, and that this 
trend holds 
for various compounds in 
the cuprate family\cite{sakakibara}.  
Namely, the strength of 
the one-band character dominates the superconductivity within this family.   

Given this background, the purpose of the present paper 
is to compare two cases: 
a multi-band model in which a flat band coexists with 
a dispersive one\cite{kobayashi}, and a single-band case in which 
a portion of the band dispersion is flat\cite{sayyad}.  
For the single-orbital multi-band system, 
the existence of the second band can actually induce the 
superconductivity especially when the second band is 
close to, but away from, the Fermi energy.  
For the single-orbital one-band system 
with a flat portion in the dispersion, 
SC can also be induced, in a manner very sensitive to 
the position of the Fermi energy.  
We shall compare these to give some hints for 
various factors dominating the flat-band SC.

\begin{figure}[h]
\begin{center}
  \includegraphics[width=0.45\textwidth]{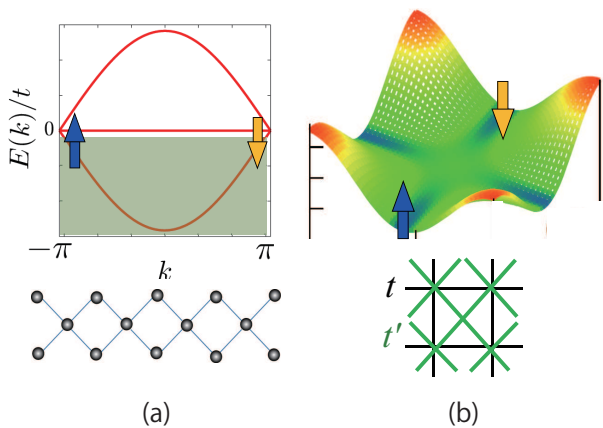}
\caption{(a) Diamond chain, which is a 
quasi-1D flat-band model, and its dispersion.  
(b) $t$-$t'$ model with the second-neighbour 
hopping (blue lines) $t'\simeq -0.5t$ , and its dispersion.  Arrows represent typical position of Cooper pairs when the 
Fermi energy is just above or below the flat part.}
\label{fig:1}    
\end{center}   
\end{figure}

\section{Multi-band systems containing flat bands}
\label{sec:1}
As a simplest possible one-dimensional flat-band model 
that contains a flat band, 
we take the diamond chain, where diamonds are connected 
into a chain [Fig.1(a)].  This model is intimately related 
with the narrow-wide band model considered by 
Kuroki and coworkers\cite{narrowwide}.  
If we consider the repulsive Hubbard model on such lattices, 
the basic idea is: when the Cooper pairs are formed 
on the dispersive band, there exist quantum mechanical 
{\it virtual pair-scattering} processes in which pairs are scattered 
between the dispersive and flat bands. 
This 
should especially be important when the flat band is 
``incipient" (i.e., away from, but close to, the Fermi energy).  
We have employed DMRG (density-matrix renormalisation group), since the system is quasi-1D, and have shown 
that: (i) we do have enhanced pairing when the flat band is incipient for intermediate repulsion $U \simeq 4t$ 
($t$: nearest-neighbour one-electron hopping), 
where the pair is spin-singlet and formed across the 
outer sites, and (ii) in that 
regime we have to take unusually large number of 
states, $m \sim 1500$ in DMRG for convergence, 
which signifies an anomalously large {\it entanglement}\cite{kobayashi}.  
In the phase diagram against band filling (Fig.2), 
the superconductivity (SC) sits just below the 
topological insulator (TI) that occurs when the dispersive 
band is just completely filled and the flat band is 
just empty.  TI is detected from entanglement 
spectra and topological edge states, and the 
situation is similar to the TI in the celebrated Haldane's 
$S=1$ antiferromagnetic chain.  

If we actually look at the pair correlation function 
against 
real space\cite{kobayashi_added} in Fig.\ref{fig:2added},
we can see the following: 
(i) The dominant (longest-tailed) 
correlation is 
between the pairs, each of which comprises 
top and bottom sites in the diamond 
chain (blue curve in Fig.\ref{fig:2added}), 
while the subdominant correlation is 
between a pair along $y$ (green).  
In classifying pairing symmetries, 
it is customary in the ladder physics to call 
a pair ``d-wave" when the 
correlation between a pair along $x$ direction and 
another along $y$ has negative values\cite{noack94}.  
In this sense, the diamond chain also has a ``d-wave", 
which is expected to tend to the d-wave in two dimensions.  
If we look at the pair correlation at distances $1 \leq r \leq 6$, 
the functions are seen to oscillate wildly, which should 
indicate some structure extended in real space. 
(ii) If we compare the diamond chain 
with the ordinary ladder system, the pair correlation function 
in the latter obtained with the quantum Monte Carlo (QMC) 
shows a long-tailed behaviour, with some wiggles related 
with the Fermi point effect involving the Fermi
wavenumber $k_F$\cite{kuroki96}.  
By contrast, the present result for the diamond chain has 
a smooth behaviour at large distances, which should be 
an effect due to the flat band.  

Recently, Matsumoto et al. have shown for 
various flat-band models that we do have a 
general trend for enhanced SC as the Fermi 
energy $E_F$ approaches the flat-band energy, with a 
sharp dip when  $E_F$ is too close to the flat band\cite{matsumoto}.  
The width of the dip depends on the Hubbard repulsion $U$, 
the degree of warping of the flat band due to many-body effects, and 
also the lattice structure, which is considered to be 
related with the self-energy effect in the flat-band 
system.  Incidentally, while the terminology ``incipient" is also 
used in the community of the iron-based FeSe 
superconductor for the incipient $s_{\pm}$ 
pairing involving the hole band below $E_F$, the 
concept of the incipient situation was originally introduced 
by Ref.\cite{narrowwide}, and we have indeed a drastic 
effect when we have a flat band instead of dispersive bands.  
We can also extend the flat-band models to two-dimensional lattices, where we can make a flat band pierce a dispersive one\cite{misumi}.  
As for candidates for realising the flat-band models, 
they include 
a mineral azurite\cite{azurite}, 
``hidden ladders" 
in Ruddlesden-Popper compounds such as Sr$_3$Mo$_2$O$_7$
\cite{ogura} 
or herbertsmithite.

\begin{figure}[ht]
\begin{center}
  \includegraphics[width=0.5\textwidth]{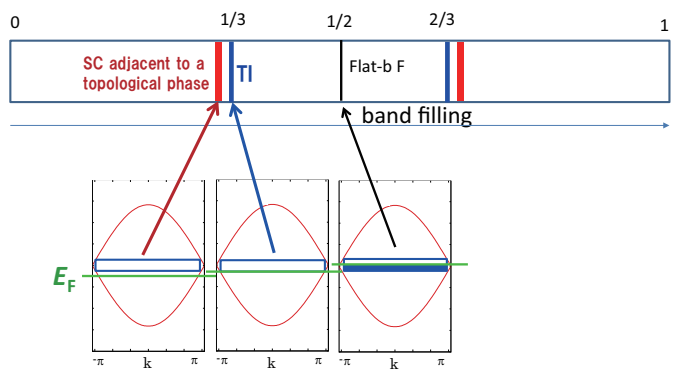}
\caption{Schematic phase diagram against the band 
filling for the diamond chain obtained with DMRG.  
Bottom panels schematically show the band filling.}
\label{fig:2}     
\end{center}  
\end{figure}

\begin{figure}[h]
\begin{center}
  \includegraphics[width=0.48\textwidth]{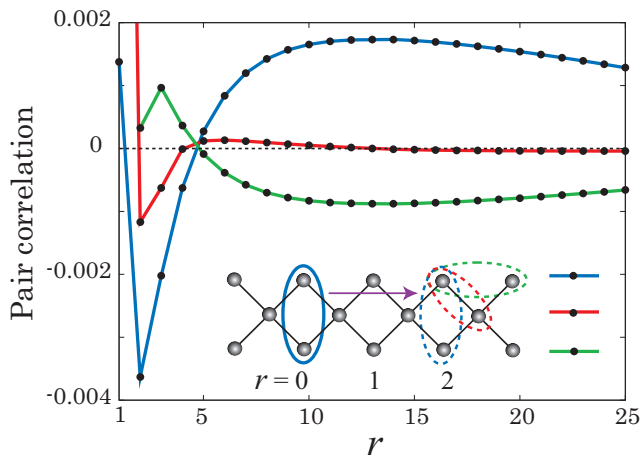}
\caption{DMRG result for the pair correlation function against 
real-space distance, $r$, between various spin-singlet pairs as 
colour-coded.}
\label{fig:2added}     
\end{center}  
\end{figure}

\section{Partially-flat one-band superconductivity}
\label{sec:2}

Having looked at a multi-band, quasi-1D model, let us move 
on to partially flat-band models that are one-band 
and two-dimensional (2D)\cite{sayyad}.  
A starting point is that, even within one-band models, 
we can have a flat {\it portion} coexisting with 
dispersive portions in the band structure [Fig.1(b)].  
An important question then is: can we still have 
flat-band superconductivity, and, 
for 2D systems, how would be the 
pair-scattering channels in 2D?  
For such a model, Huang et al.\cite{vaezi} have 
studied superconductivity for attractive interaction 
$U$ and Mott insulation for repulsive $U$ in the 
Hubbard model with the determinantal quantum Monte Carlo 
(DQMC) method.  They have detected an electron correlation 
effect for less-than-half-filled cases unlike in the 
ordinary bands.

\begin{figure*}[ht]
\begin{center}
  \includegraphics[width=0.7\textwidth]{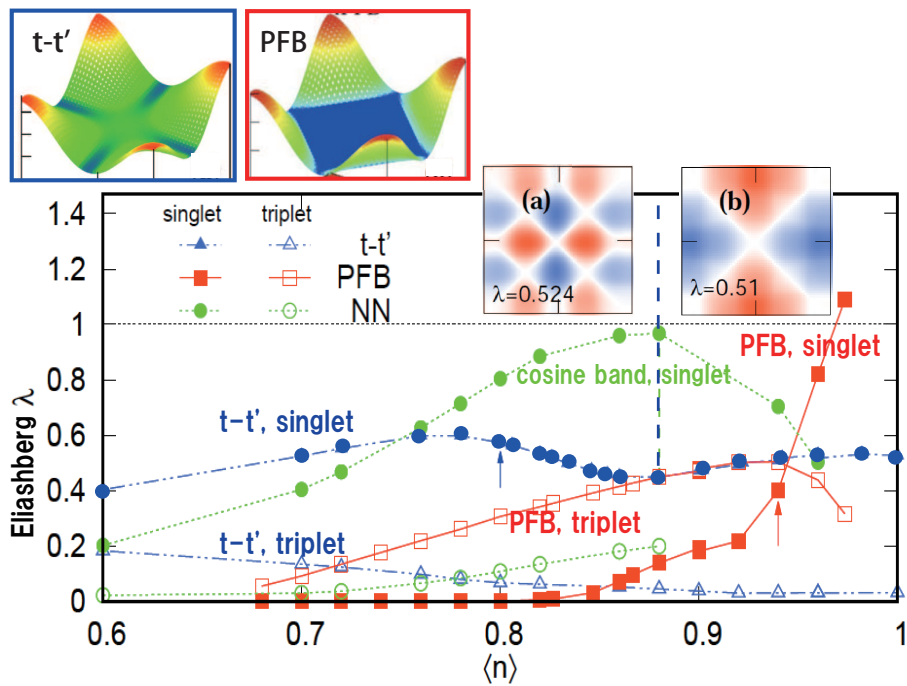}
\caption{For $t$-$t'$ model (top left inset) and 
truncated model (right), the eigenvalue, $\lambda$, of the Eliashberg equation versus the band filling $\langle n \rangle$ is shown 
for the singlet (filled symbols) and triplet (empty) pairings for $t$-$t'$ (triangles) and 
truncated (squares), and cosine-band (circles) models for $U = 3, 1/(k_BT) = 33$.  A 
vertical line for the $t$-$t'$ singlet indicates a change in the pairing
symmetry as shown by panels next to it.}
\label{fig:3}   
\end{center}
\end{figure*}

Here, our interest is 
superconductivity for repulsive interaction.  For that 
we take two models (Fig.\ref{fig:3}): one is 
a $t$-$t'$ model on a square lattice 
with a second-neighbour hopping $t'\simeq -0.5t$ to have 
a flat portion along $k_x \simeq 0$ and $k_y \simeq 0$.  
The other model is a partially flat band (PFB) with a truncated flat bottom.  
We employ FLEX+DMFT method, where we combine 
the fluctuation-exchange approximation and the dynamical 
mean-field theory\cite{kitatani}.   
The result for the double occupancy shows that the 
correlation effect indeed emerges well below half-filling 
and even for small repulsive $U$ in both models.  
If we look at 
the momentum distribution, we can see that electrons are 
crammed into the flat portion, which should be 
the reason for the early onset of the double occupancy.
The result for the spin susceptibility, $\chi_S$, 
exhibits that, for the $t$-$t'$ model, we have large amplitude of $\chi_S$ 
along some ridges in $k$-space that becomes wider 
as we approach the half-filling, or, for the truncated 
model, wide plateaux for $\chi_S$ that move from 
around the $\Gamma$ point towards the AF points, $(\pm\pi, \pm\pi)$, 
as we approach the half-filling. 
If we turn to the eigenvalue, $\lambda$, 
of the Eliashberg equation in Fig.\ref{fig:3}, in the $t$-$t'$ model, 
the spin-singlet pairing dominates, where $\lambda$ 
exhibits a {\it double-dome} structure.  We can 
identify its origin: the peak on the smaller-filling 
side represents a complicated gap function that 
have a larger number of nodes than in the usual 
d-wave, while the peak on the larger 
filling represents a pairing close to the d-wave. 
For the truncated 
model, on the other hand, triplet pairing dominates over an unusually 
wide filling region with a p-wave-like gap function, 
which is sharply taken over, as the half-filling is approached, by singlet pairing with a gap function 
having a larger number of nodes than in the d-wave. 
If we look at the pairing in real space, the case 
of large numbers of nodes exhibits unusually extended 
pairing in real space.  
We have also detected non-Fermi-liquid properties 
from the frequency dependence of the self energy\cite{sayyad}.

\begin{figure*}[hb]
\begin{center}
  \includegraphics[width=0.85\textwidth]{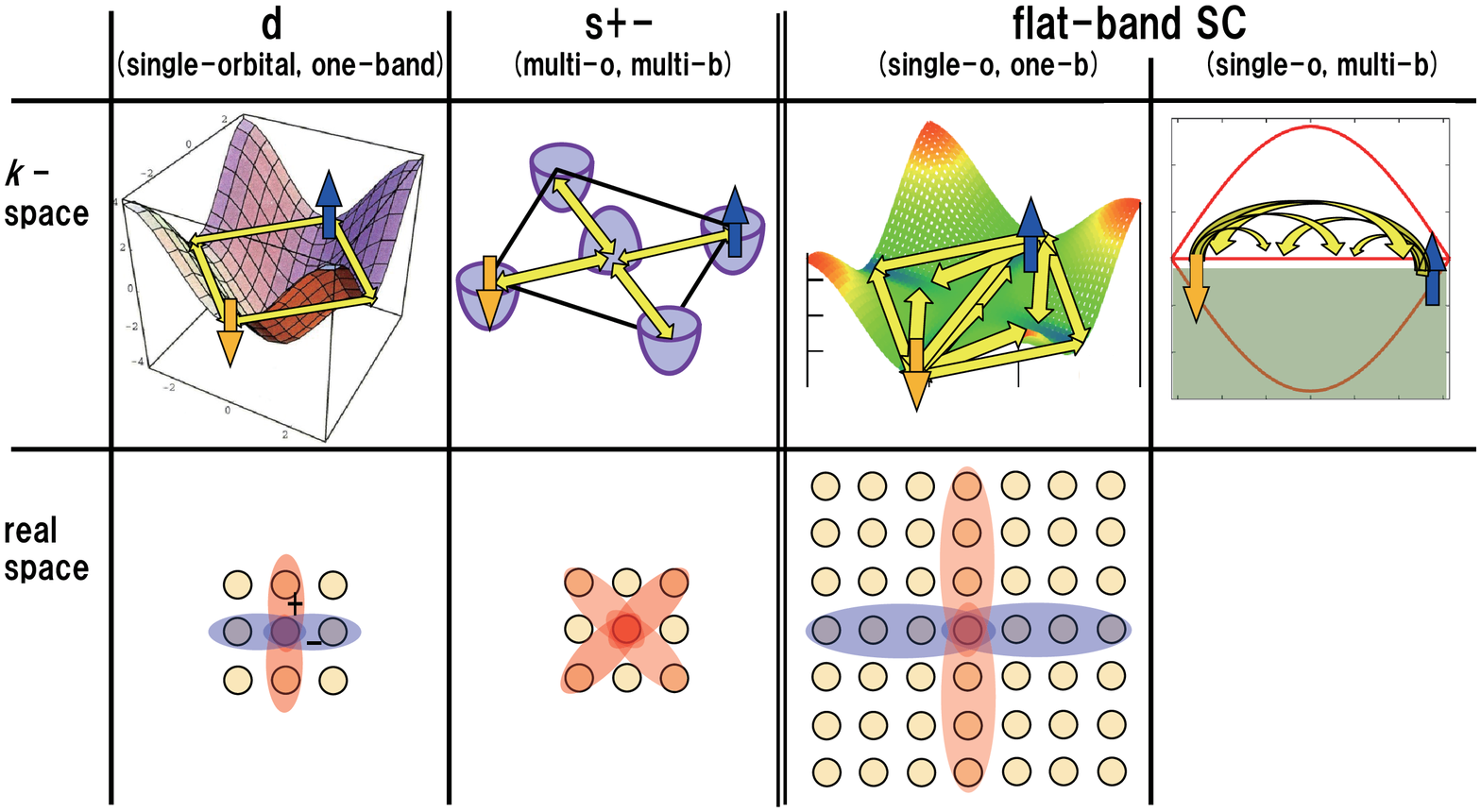}
\caption{We schematically compare ordinary single-orbital, one-band case (here for a d-wave SC; leftmost column) and
multi-orbital, multi-band case (here for $s_{\pm}$; second column from left), 
where the nesting vectors (yellow arrows) 
connecting  the specific ``hot spots" 
designate how pairs (blue and orange arrows) hop. These are contrasted with
flat-band systems for single-orbital, one-band case (second from right) and single-orbital, multi-band case (rightmost), where yellow arrows represent pair-scattering channels. The top
row depicts $k$-space, while the bottom row displays pairs in real space.}
\label{fig:4}   
\end{center}
\end{figure*}

All these are considered 
to come from the flat portions in the band dispersion.  
As summarized in Fig.\ref{fig:4} for electron-mechanism superconductivity, usually, we have well-defined 
nesting vectors that connect ``hot spots", 
which are the anti-nodal 
regions in single-orbital, one-band systems as 
in the d-wave in the cuprates, or the 
electron and hole pockets in multi-orbital, 
multi-band systems as in the $s_{\pm}$-wave 
in the iron-based\cite{hosonoKuroki}.   By contrast, 
in partially flat-band systems 
we have a {\it bunch} of pair-scattering 
channels, which should be the cause for the peculiar 
spin structures and the ensuing gap functions there.

\section{Discussions}

Can flat bands really favour SC?  We can raise several points 
on this.  

(i) {\it Dimensionality}: Electron-mechanism SC employing spin-fluctuation mediated 
pairing usually uses well-defined spin structure such as 
AF fluctuations, and this results in 
specific and compact regions in $k$-space in which 
the pairing interaction is strong.  From the phase volume arguments, 
we can then show that quasi-2D (layered) systems are much more favourable 
for such SC than in 3D systems\cite{arita}.  
This is also consistent with the empirical fact that most of the 
new superconductors have layered structures.  By contrast, 
the flat-band systems have much wider momentum regions for large spin 
structure, $\chi_S$.  Another factor in the spin-fluctuation mediated 
pairing interaction is Green's function $G$ involved in the 
Eliashberg equation, and $G$ too 
exhibits wide regions for the flat bands.  
How about the structure of the gap function?  
In a multi-band case of the narrow-wide band model\cite{narrowwide}, 
we have an ``$s_{\pm}$ wave" between the flat and 
dispersive bands, where each band has more or less 
homogeneous amplitude in k-space, which comes from 
a featureless $\chi_S$.  The gap function in the 
diamond lattice also has an extended structure 
as seen from the pair-pair correlation function (Fig.\ref{fig:2added}) 
that is long-tailed and long-ranged structure in 
real space\cite{kobayashi}.  So we can expect that 
3D systems may be as good as 2D systems in the flat-band SC.

(ii) {\it Vertex corrections}: In general, the size of $T_C$ in SC arising from 
electron-electron repulsion is shown to involve the vertex correction 
in the pair scattering\cite{kitataniDGA}, which is identified 
as the main 
reason why $T_C$ is two orders of magnitude lower than the 
electronic energy.  Thus the vertex correction in the flat-band 
systems is an interesting future problem.  
Incidentally, flat bands have also been 
discussed where the many-body renormalised mass 
(which enters in the Fermi liquid theory) is heavy,\cite{khodel}  
whereas we consider here non-interacting bands 
that are dispersionless.  
A heavy renormalised mass (occurring e.g. when
the Fermi energy is right at the one-electron flat
band that makes the self-energy correction large), 
which is contrasted with the present case of {\it incipient} flat 
band or portion that can work favourably for 
superconductivity.


(iii) {\it Strong-coupling limit}: 
In the 
strong-coupling limit, the Hubbard model is converted into 
a Heisenberg spin model.  It has been shown that a kind of 
Creutz model (with cross-linked interactions) in 2D 
can accommodate a supersolid phase where superfluid and 
density wave coexist\cite{murakami}.  

(iv) {\it Superfluid weight}: 
T\"{o}rm\"{a} and coworkers 
have shown, for the attractive electron-electron interaction, 
that superfluidity in topological flat bands 
has a superfluid weight lower-bounded by the topological 
number\cite{torma}.  
So the question of what happens for 
repulsive interactions will be another interesting future problem.  

If we summarize the comparison for the 
flat-band systems between the multi-band 
and one-band cases, a similarity is 
that superconductivity can be induced in both cases 
when the flat band or portion is incipient.  
The differences are mainly the different 
gap function structures between the two cases, 
which is caused by different spin structures 
as dictated by the band dispersion, and then 
results in difference in the pairing symmetries.  
In the multi-band diamond chain, the pairing exploits 
large entanglement arising from the topological 
nature of the flat band.  Partially-flat one-band 
systems may accommodate something related 
with topological states.  

The group velocity vanishes at van Hove singularities, 
and there are literatures discussing a possibility 
of topological superconductivity involving the 
van Hove singularities\cite{volovik1994,yudin2014,liu2018}.  
In the flat-band systems the group velocity vanishes 
over a finite area rather than at a point.  There, we have observed a transition 
between p-wave and d-wave\cite{sayyad}.  It is generally recognized 
that the boundary region between different 
pairing symmetries are a good place for looking for 
topological superconductivity with broken 
time-reversal symmetry\cite{fernandes2013}, 
so there may be a possibility of topological SC in 
partially flat-band systems.  
Also, partially flat bands remind us of the band 
structure of the twisted bilayer graphene 
for which SC was discovered, 
and first-principles calculations show 
partial flatness\cite{TBG,Guinea,Vishwanath}.  
However, this material 
involves various complications such as a multi-band 
character, so the present one-band model will not 
apply directly.  
If we go over to three-dimensional systems, Akashi has recently 
shown that ``saddle loops" (an extension of van Hove saddle 
points) can occur from a general standpoint\cite{akashi}, which 
may be utilised for band structure engineering.

In a broader context than the flat-band physics, studies of 
superconductivity in two-band systems has a long history, 
basically starting from Suhl-Kondo mechanism.  For {\it repulsively} 
interacting two-band systems, Kuroki and the present author have 
investigated superconductivity with QMC\cite{kurokiQMC} 
and the bosonisation\cite{kurokiBosonisation}.  
For the cuprates specifically, effects of hydrostatic and uniaxial 
pressure for multi-band models\cite{sakakibara} or strain control of multigap superconductors \cite{agrestini} have  been discussed.  
For {\it attractively} interacting two-band systems, the effects of 
the second band have been studied with the Nozi\`{e}res-Schmitt-Rink 
formalism\cite{tajima}, and the multiband Suhl-Matthias-Walker Hamiltonian\cite{bussmann}.  
The effect of Lifshitz transitions has also been examined\cite{bianconiJarlborg} in terms of 
the Fano resonance between the flat and dispersive bands in the BCS-BEC crossover regime\cite{kagan}.
Thus a future problem is how these would apply to different realisations of flat-band cases.

\begin{acknowledgements}
The author wishes to thank the collaborators of 
Refs.\cite{kobayashi,sayyad} that are described in the present 
article.  He also 
acknowledges a support from
the ImPACT Program of the Council for Science, Technology
and Innovation, Cabinet Office, Government of Japan (Grant
No. 2015-PM12-05-01) from JST, 
JSPS KAKENHI Grant Nos. JP26247057, 17H06138, and
CREST ``Topology" project from JST.
\end{acknowledgements}

%
\section*{Conflict of interest}
The authors declare that they have no conflict of interest.


\begin{thebibliography}{}
%

\bibitem{sakakibara} H. Sakakibara et al., Phys. Rev. Lett. {\bf 105}, 057003 (2010); Phys. Rev. B {\bf 85}, 064501 (2012); {\bf 86}, 134520 (2012); {\bf 89}, 224505 (2014).

\bibitem{kobayashi} K. Kobayashi, M. Okumura, S. Yamada, M. Machida, H. Aoki, Phys. Rev. B {\bf 94}, 214501 (2016).

\bibitem{sayyad} S. Sayyad, E. W. Huang, M. Kitatani, M.-S. Vaezi, Z. Nussinov, A. Vaezi and H. Aoki, Phys. Rev. B {\bf 101}, 014501 (2020).

\bibitem{narrowwide} K. Kuroki, T. Higashida, and R. Arita, Phys. Rev. B {\bf 72}, 212509 (2005).

\bibitem{kobayashi_added} K. Kobayashi, M. Okumura, S. Yamada, M. Machida, and H. Aoki, unpublished.

\bibitem{noack94} R.M. Noack, S.R. White and D.J.Scalapino, Phys. Rev. Lett. {\bf 73}, 882 (1994).  In the Tomonaga-Luttinger theory, 
a pairing is called d-wave-like when the 
pair wave function has opposite signs between 
the bonding and antibonding bands, as in 
L. Balents and M.P.A. Fisher, Phys. Rev. B {\bf 53}, 12133 (1996). 

\bibitem{kuroki96} K. Kuroki, T. Kimura and H. Aoki, Phys. Rev. B {\bf 54}, R15641 (1996).

\bibitem{matsumoto} K. Matsumoto, D. Ogura, and K. Kuroki, Phys. Rev. B {\bf 97}, 014516 (2018).  

\bibitem{misumi} T. Misumi and H. Aoki, Phys. Rev. B {\bf 96}, 155137 (2017).

\bibitem{azurite} H. Kikuchi et al., Phys. Rev. Lett. {\bf 94}, 227201 (2005); H. Jeschke et al., Phys. Rev. Lett. {\bf 106}, 217201 (2011).

\bibitem{ogura} D. Ogura, H. Aoki and K. Kuroki, Phys. Rev. B {\bf 96}, 184513 (2017).

\bibitem{vaezi} E. W. Huang, M.-S. Vaezi, Z. Nussinov, and A. Vaezi, Phys. Rev. B {\bf 99}, 235128 (2019).

\bibitem{kitatani} M. Kitatani, N. Tsuji and H. Aoki, Phys. Rev. B  {\bf 92}, 085104 (2015).

\bibitem{hosonoKuroki} H. Hosono and K. Kuroki, Physica C {\bf 514}, 399 (2015).

\bibitem{arita} P. Monthoux and G. G. Lonzarich, Phys. Rev. B {\bf 59}, 14598 (1999); R. Arita, K. Kuroki, and H. Aoki, Phys. Rev. B {\bf 60}, 14585 (1999).

\bibitem{kitataniDGA} M. Kitatani, T. Sch\"{a}fer, H. Aoki and K. Held, Phys. Rev. B {\bf 99}, 041115(R) (2019).

\bibitem{khodel} V.A. Khodel and V.R. Shaginyan, 
JETP Lett. {\bf 51}, 553 (1990).

\bibitem{murakami} Y. Murakami, T. Oka and H. Aoki, Phys. Rev. B {\bf 88}, 224404 (2013).

\bibitem{torma} S. Peotta and P. T\"{o}rm\"{a}, Nature Commun. {\bf 6}, 8944 
(2015); M. Tovmasyan et al., Phys. Rev. B {\bf 94}, 245149 (2016); {\bf 98}, 134513 (2018); 
A. Julku et al, Phys. Rev. B {\bf 101}, 060505(R) (2020).

\bibitem{volovik1994} G.E. Volovik, JETP Lett. {\bf 59}, 830 (1994).

\bibitem{yudin2014} D. Yudin, D. Hirschmeier, H. Hafermann, O. Eriksson, A.I. Lichtenstein, M.I. Katsnelson, Phys. Rev. Lett. {\bf 112}, 070403 (2014).

\bibitem{liu2018} C. C. Liu et al., Phys. Rev. Lett. {\bf 121}, 217001 (2018).

\bibitem{fernandes2013} F. M. Fernandes and A. J. Millis, Phys. Rev. Lett. 
{\bf 111}, 127001 (2013); F. Ahn et al., Phys. Rev. B {\bf 89}, 144513 (2014).

\bibitem{TBG} Y. Cao, V. Fatemi, S. Fang, K. Watanabe, T. Taniguchi, E. Kaxiras, and P. Jarillo-Herrero, Nature {\bf 556}, 43 (2018); 
X. Lu et al, Nature {\bf 574}, 653 (2019).

\bibitem{Guinea} F. Guinea and N.R. Walet, Proc. Nat. Acad. Sci. {\bf 115}, 
13174 (2018).

\bibitem{Vishwanath} H.C. Po et al, Phys. Rev. B {\bf 99}, 195455 (2019).

\bibitem{akashi} R. Akashi, Phys. Rev. B {\bf 101}, 075126 (2020).

\bibitem{kurokiQMC} K. Kuroki and H. Aoki, Phys. Rev. B {\bf 42}, 2125 (1990); Phys. Rev. Lett. {\bf 69}, 3820 (1992); Phys. Rev. B {\bf 48}, 7598 (1993).

\bibitem{kurokiBosonisation} K. Kuroki and H. Aoki, 
Phys. Rev. Lett. {\bf 72}, 2947 (1994).

\bibitem{agrestini} S. Agrestini et al, J. Phys. A {\bf 36}, 9133 (2003). 

\bibitem{tajima} H. Tajima et al, Physical Review B {\bf 99}, 180503 (2019).

\bibitem{bussmann} A. Bussmann-Holder et al, Condensed Matter {\bf 2}, 24 (2017).

\bibitem{bianconiJarlborg}  A. Bianconi, Journal of Superconductivity, {\bf 18}, 625 (2005);  D. Innocenti et al.  Superconductor Science and Technology, {\bf 24}, 015012 (2010); A. Bianconi and T. Jarlborg, EPL (Europhysics Letters) {\bf 112}, 37001 (2015);  T. Jarlborg, A. Bianconi,  Scientific Reports, {\bf 6}, 24816, (2016); M.V. Mazziotti, et al. EPL (Europhysics Letters), {\bf 118}, 37003 (2017). 

\bibitem{kagan} M.Y. Kagan and A. Bianconi, Condensed Matter {\bf 4}, 51 (2019).
\end{thebibliography}


\end{document}